\documentclass[twocolumn,english,pra,showpacs,preprintnumbers,amsmath,amssymb]{revtex4-1}
\usepackage{graphicx}
\usepackage{amssymb}

\usepackage{babel}

\begin{document}


\title{Disentanglement in Bipartite Continuous-Variable Systems}

\author{F. A. S. Barbosa$^1$, A. J. de Faria$^2$, A. S. Coelho$^1$, K. N. Cassemiro$^{1,3}$, A. S. Villar$^{1,3,4}$,
P. Nussenzveig$^1$, and M. Martinelli$^1$ }

\affiliation{$^1$ Instituto de F\'\i sica, Universidade de S\~ao Paulo, P.O. Box 66318, 05315-970 S\~ao Paulo, Brazil. \\
$^2$ Instituto de Ci\^encia e Tecnologia, Universidade Federal de Alfenas, 37715-400 Po\c{c}os de Caldas, MG, Brazil \\
$^3$ Max Planck Institute for the Science of Light, 91058 Erlangen, Germany. \\
$^4$ Lehrstuhl fuer Optik, Universitaet Erlangen-Nuernberg, 91058 Erlangen, Germany.} 

\email{mmartine@if.usp.br}


\date{\today}


\begin{abstract}
Entanglement in bipartite continuous-variable systems is investigated in the presence of 
partial losses, such as those introduced by a realistic quantum communication channel, e.g. by 
propagation in an optical fiber. We find that entanglement can vanish completely for partial 
losses, in a situation reminiscent of so-called entanglement sudden death. Even states with 
extreme squeezing may become separable after propagation in lossy channels. Having 
in mind the potential applications of such entangled light beams to optical communications, 
we investigate the conditions under which entanglement can survive for all partial losses. 
Different loss scenarios are examined and we derive criteria to test the robustness of 
entangled states.  These criteria are necessary and sufficient for Gaussian states. Our 
study provides a framework to investigate the robustness of continuous-variable 
entanglement in more complex multipartite systems. 

\end{abstract}



\maketitle

\section{Introduction}

The dynamics of open quantum systems leads in general to a degradation of 
key quantum features, such as coherence and entanglement. Since entanglement 
is considered to be an important resource for applications in quantum information, 
its degradation may seriously hinder the envisioned protocols. Careful analyses of 
environment-induced loss of entanglement are thus important steps in quantum 
information science. In the discrete-variable scenario, studies of 2-qubit systems 
have shown that entanglement can be completely lost after a finite time of interaction 
with the environment, an effect now mostly known as Entanglement Sudden Death 
(ESD)~\cite{Eberly,Davidovich}. Quantum information can also be conveyed, stored, 
and processed by continuous-variable (CV) systems. Bright beams of light can be 
described by means of CV field quadratures and are natural conveyors of quantum 
information. Unavoidable transmission loss is the fiercest enemy for quantum 
communications. It has recently been observed that losses may lead to complete 
disentanglement in Gaussian CV systems~\cite{CoelhoScience,Barbosa2010}. 
This phenomenon is a partial-loss analog of the finite-time disentanglement observed 
in qubit systems.

The simplest CV systems one can consider are those described by Gaussian statistics. 
Gaussian states are indeed well studied~\cite{braunsteinvanloockRMP} and fairly well 
characterized. For instance, there exist necessary and sufficient criteria for Gaussian-state 
entanglement of up to $1\times N$ systems (in which one subsystem is collectively 
entangled to $N$ other subsystems)~\cite{Simon,wernerwolf}. In spite of all this 
knowledge, the sensitivity of entanglement to the interaction with the environment is 
still not completely mapped. As experimentally observed by Coelho {\it et al.}~\cite{CoelhoScience} 
and by Barbosa {\it et al.}~\cite{Barbosa2010}, some Gaussian states become separable for 
partial losses while others remain entangled. What distinguishes one class of states from 
the other? Are there only two classes of such states? Is it sufficient to produce states with 
a large degree of squeezing in order to avoid disentanglement? Is there any strategy 
involving local operations to protect states against disentanglement?

In this paper, we extend the treatment of ref.~\cite{Barbosa2010} and provide answers to 
some of these questions. We theoretically analyze the conditions leading to CV disentanglement 
in the simplest case of bipartite systems. In the framework of open system dynamics, the effect 
of a lossy channel (without any added noise) is equivalent to the interaction with a reservoir at zero 
temperature. The property of entanglement resilience to losses will be referred to as `robustness'. 
Entanglement robustness is assessed by entanglement criteria previously derived by other authors. 
For general CV states, these criteria provide sufficient conditions for the robustness of bipartite 
systems. Necessary and sufficient entanglement criteria for Gaussian states lead to necessary 
and sufficient conditions for entanglement robustness upon propagation in lossy channels. 
Entanglement of CV Gaussian states may be created by a number of different strategies 
such as, for instance, passive operations on initially squeezed states~\cite{wolfeisertplenio2003}. 
We shall not discuss these in detail here, but take for granted initially entangled states. 

\begin{figure}[ht] 
\includegraphics[width=8.5cm]{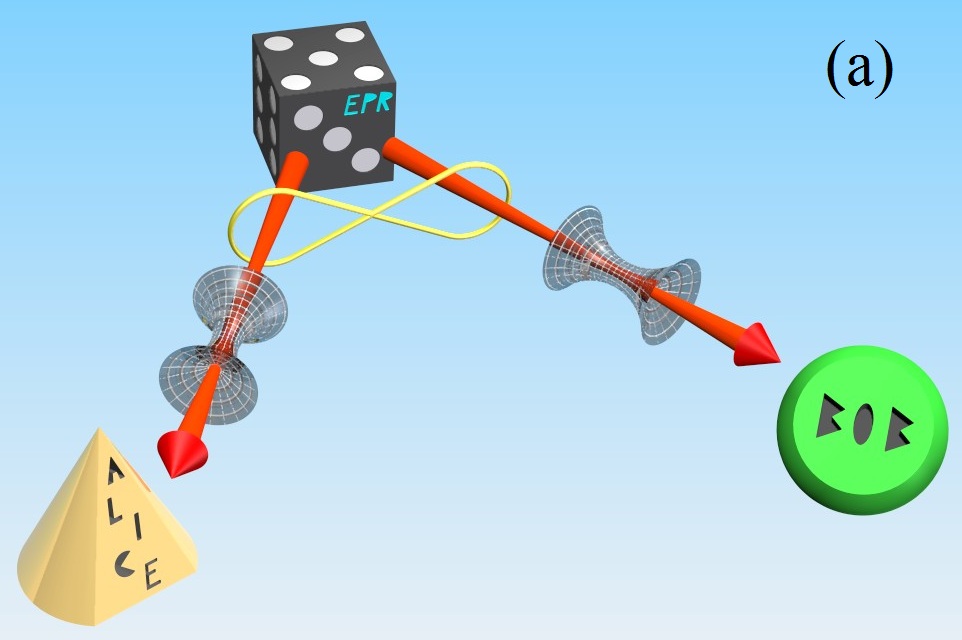} \includegraphics[width=8.5cm]{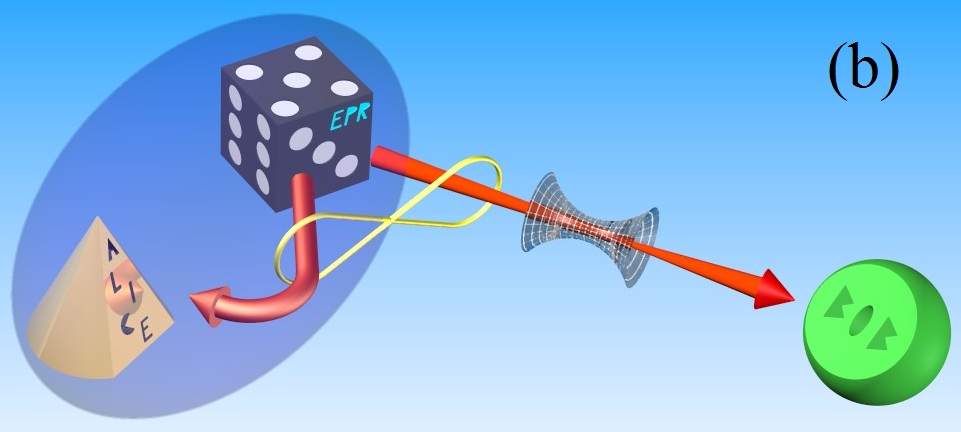}
\caption{(Color online) (a) {\em Dual-Channel Losses}: An entangled quantum state is distributed to 
two parties, Alice and Bob, over two lossy quantum channels; (b) {\em Single-Channel 
Losses}: Alice holds the quantum state generator and only distributes 
one entangled mode to Bob over a single lossy quantum channel.}
\label{fig:singlevsdualchannel}
\end{figure}

A thorough investigation reveals the possibility of distinct entanglement dynamics as 
losses are imposed on the subsystems. We consider realistic scenarios, as depicted 
in Fig.~\ref{fig:singlevsdualchannel}. A bipartite entangled state is the quantum resource 
of interest. It can be distributed to two parties who wish to communicate, as in 
Fig.~\ref{fig:singlevsdualchannel} (a), in a scenario which we refer to as a dual-channel 
communication scheme. Another possibility would be that one of the parties holds 
the quantum state generator and only one mode needs to propagate through a 
lossy quantum channel, as in Fig.~\ref{fig:singlevsdualchannel} (b). We refer to this 
situation as a single-channel scheme. One could surmise, in principle, that it is equivalent 
to concentrate losses in a single channel or split them among two channels. If 
our channels are optical fibers, losses increase exponentially with the propagation 
distance. Thus, one could think that propagation in a single fiber over a certain distance 
would have the same effect as propagation of both modes, each in one fiber, over half 
the distance (which would result in the same overall losses). This is not correct: for certain states, 
one could propagate one of the modes over an infinite distance in a single 
lossy channel without losing entanglement, whereas entanglement would disappear after 
a finite propagation distance if both modes were to suffer losses. 

These different scenarios lead to the introduction of a formal classification, consisting 
of three robustness classes. On one extreme, the entanglement of {\it fully robust} states vanishes 
only for total attenuation of either beam. On the opposite extreme, {\it fragile states} become 
separable for partial attenuations on either beam or a combination of both. An intermediate 
class of {\it partially robust} states shows either robustness or fragility depending on the way 
losses are introduced. Thus, imposing losses on one field may be less harmful in a quantum 
communication system than distributing both beams over two lossy channels. Furthermore, 
we show that even states with very strong squeezing (e.g. amplitude difference squeezing, 
as in twin beams produced by an above-threshold OPO) can disentangle for partial losses. 
A moderate excess noise, commonly encountered in existing experiments, suffices for this. 
In addition, one could speculate that pure states would necessarily be robust. We provide 
an example of a pure state that disentangles for partial losses as well. 

The paper is organized as follows. In Section~\ref{lossy} we establish notation and the 
basic reservoir model (the environment). In Section~\ref{robust} a sufficient criterion to 
determine the robustness of the entangled state is demonstrated. In Section~\ref{realrobust} 
we extend the robustness criterion, resulting in a necessary and sufficient robustness 
condition for all Gaussian bipartite states. The different classes of entanglement robustness 
against losses in each channel are defined in Section~\ref{class}. In Section~\ref{examples} 
we examine particular quantum states commonly treated in the literature.  A final Section~\ref{conclusion} 
is focused on the main physical results and implications of our findings. 

\section{Entanglement and ESD in Lossy Gaussian Channels \label{lossy}}

The quantum properties of Gaussian states are completely
characterized by the second order moments of the appropriate observables.
The choice of observables depends on the system under consideration.
In the case of the electromagnetic field, a complete description can be 
given in terms of orthogonal field quadratures. We will consider
the amplitude and phase quadratures, respectively
written as $\hat{p}_{j}=(\hat{a}_{j}^{\dag}+\hat{a}_{j})$ and
$\hat{q}_{j}=i(\hat{a}_{j}^{\dag}-\hat{a}_{j})$ in terms of the
field annihilation $\hat{a}_{j}$ and creation $\hat{a}^\dag_{j}$
operators. The indices $j=1,2$ stand for the two field modes of our bipartite system.
The quadrature operators obey the commutation relation $[\hat{p}_{j},\hat{q}_{j}]=2i$, 
from which we obtain an uncertainty product lower bound of one. The standard 
quantum level (SQL) is thus equal to one, representing the noise power present in 
the quadrature fluctuations of a coherent state.

It is useful to organize the second order moments in the form of a $4\times4$ covariance matrix $V$.
Its entries are the averages of the symmetric products of quadrature fluctuation operators
\begin{equation}
V=\frac{1}{2}\left\langle\delta\hat\xi\delta\hat\xi^T+(\delta\hat\xi\delta\hat\xi^T)^T\right\rangle\,,
\label{covariance}
\end{equation}
where $\hat\xi=( \hat q_1, \hat p_1, \hat q_2 ,\hat p_2 )^T$ is the column vector of quadrature 
operators, and $\delta\hat\xi=\hat\xi-\langle\hat\xi\rangle$ are the fluctuation operators with 
zero average. Similar notation will be valid for the individual quadratures, e.g. $\delta\hat p_1$.
The noise power is proportional to the variance of the fluctuation,
denoted for a given quadrature by (e.g.)  $\Delta^2\hat p_1=\langle(\delta\hat p_1)^2\rangle$.
The Heisenberg uncertainty relation can be expressed as~\cite{Simon94,Simon}
\begin{eqnarray}
V+i\Omega & \geq & 0,
\label{physicality}
\\
\mathrm{where}\quad\Omega=\left[\begin{array}{cc}
J & 0\\
0 & J\end{array}\right], & \,\mbox{and}\, & J=\left[\begin{array}{cc}
0 & 1\\
-1 & 0\end{array}\right].\nonumber
\end{eqnarray}
The covariance matrix can be divided in three $2\times2$ submatrices,
from which two ($A_j$) represent the reduced covariance matrices of the individual subsystems
and one ($C$) expresses the correlations between the subsystems
\begin{eqnarray}
 &  & V=\left(\begin{array}{cc}
A_1 & C\\
C^T & A_2\end{array}\right).
\label{Covar}
\end{eqnarray}
The correlations originate from both classical and quantum backgrounds,
and cannot be directly associated to entanglement without considering the
properties of each subsystem. As we will see, the occurrence of ESD is 
related to the presence of uncorrelated noise in the system, normally in 
the form of unbalanced or insufficient correlations between different subsystems or quadratures.

For bipartite Gaussian states, there exist necessary and sufficient entanglement 
criteria~\cite{Simon,DGCZ}. These criteria are the basis for our assessment of 
entanglement robustness.

First, we need to adopt a model for the quantum channel. Here
we consider the realistic case of a lossy bosonic channel, equivalent to
the attenuation of light by random scattering. 
Losses are modeled by independent beam splitters placed in the beam paths. 
Each beam splitter transformation combines one field mode with the vacuum field. 
In the absence of added noise, it can be associated to a reservoir at zero 
temperature. 

A Gaussian attenuation channel transforms the field operators according to~\cite{Holevo01,Eisert05}
\begin{equation}
\hat{a}_{j}\longrightarrow\hat{a}'_{j}=\sqrt{T_{j}}\,\hat{a}_{j}+\sqrt{1-T_{j}}\,\hat{a}_{j}^{(E)},
\label{quadsplitt}
\end{equation}
where $T_{j}$ is the beam splitter transmittance and $\hat{a}_{j}^{(E)}$
is the annihilation operator from the environment. It acts 
on the covariance matrix as 
\begin{equation}
V'=\mathcal{L}(V)=L(V-I)L+I,
\label{attTrans}
\end{equation}
where $L=\mathrm{diag}(\sqrt{T_{1}},\sqrt{T_{1}},\sqrt{T_{2}},\sqrt{T_{2}})$ is the loss matrix 
and $I$ is the $4\times4$ identity matrix.

The question we address here regards the behavior of entanglement as the covariance 
matrix undergoes the transformation of Eq.~(\ref{attTrans}).

\section{The Duan entanglement criterion and Robustness \label{robust}}

We direct our attention, in a first moment, to the entanglement criterion presented
in Ref.~\cite{DGCZ}, here referred to as the Duan criterion.
According to them, a sufficient condition for the existence
of entanglement is obtained by fulfilling the inequality
\begin{equation}
W_{D}=\Delta^2 \hat u+\Delta^2 \hat v-\left(a^2+\frac{1}{a^2}\right)<0,
\label{Duan}
\end{equation}
where
\begin{equation}
\hat{u}=\frac{1}{\sqrt{2}}\left(|a|\hat{p}_{1}-\frac{1}{a}\hat{p}_{2}\right)
\; \mbox{and} \;\;\hat{v}=\frac{1}{\sqrt{2}}\left(|a|\hat{q}_{1}+\frac{1}{a}\hat{q}_{2}\right) \;. 
\end{equation}
The $\hat{p}_i$ and $\hat{q}_i$ are quadrature operators, obeying the commutation 
relations stated above and $a$ is an arbitrary real nonzero number. The quadrature 
combinations $\hat{u}$ and $\hat{v}$ are collective operators corresponding to 
the original example of Einstein, Podolsky and Rosen (EPR)~\cite{EPR}. As such, 
they are called EPR-like collective operators.

The quantity $W_D$ can be viewed as an entanglement witness. We shall use the 
symbol `$W$' for witnesses in general. The presence of a given property is signaled by a 
negative value of the corresponding witness. As a merely sufficient criterion, no statement 
can be made if $W_D\geq0$: the state could be either separable or entangled. Nevertheless, 
the witness $W_D$ is compelling from a practical point of view because it does not require 
full knowledge of the covariance matrix, simplifying the detection of entanglement in experiments. 
The downside is its limited detection ability.

For $a=1$, entanglement can be detected by a balanced beam 
splitter transformation of the input fields followed by a measurement of squeezing in the 
two output fields~\cite{Bohr,Leuchs}. Alternatively, one can measure the quadrature variances 
$\Delta^2\hat p_i$ and $\Delta^2\hat q_i$ of each field and the cross correlations
$c_{p}=\langle\delta\hat{p}_{1}\,\delta\hat{p}_{2}\rangle$ and $c_{q}=\langle\delta\hat{q}_{1}\,\delta\hat{q}_{2}\rangle$.
The optimum choice for the parameter $a$ that minimizes $W_D$ is 
$a^{2}=\sqrt{\sigma_{2}/\sigma_{1}}$, where the $\sigma_j$ are given by 

\begin{equation}
\sigma_{j}=\Delta^2\hat{p}_{j}+\Delta^2\hat{q}_{j}-2=\mathrm{tr} A_{j}-2.
\label{ExcNoise}
\end{equation}
The sign indeterminacy in $a$ is solved by taking into account the signs of the quadrature
correlations. With these considerations, one arrives at the minimized form of the Duan criterion
\begin{equation}
W_{M}=\sigma_{1}\sigma_{2}-(c_{p}-c_{q})^{2}<0.
\label{W_M}
\end{equation}

Eq.~(\ref{W_M}) provides the first insight into the robustness of bipartite states.
The crucial fact to be observed is that the sign of $W_{M}$ is conserved by attenuations.
In fact, using Eq.~(\ref{attTrans}), the correlations transform as $c_{p}'=\sqrt{T_{1}T_{2}}\,c_{p}$
and $c_{q}'=\sqrt{T_{1}T_{2}}\,c_{q}$, while $\sigma_{j}'=T_{j}\sigma_{j}$.
The attenuation operation factorizes in the entanglement witness,
\begin{equation}
W_{M}'=T_{1}T_{2}W_{M}.
\end{equation}
Therefore, an initially entangled state satisfying Eq.~(\ref{W_M}) will not disentangle 
under partial losses. This fact was experimentally verified by Bowen 
{\it et al.}~\cite{BowenPRL2003}.

Entangled states satisfying the Duan criterion do not disentangle for partial losses imposed on any mode:
they are {\it fully robust}. Among them lie the two-mode squeezed states, a large class of states for
which both EPR-like observables are squeezed~\cite{KimbleFurusawa, Leuchs, VillarPRL}.

Since $W_{M}$ is only a sufficient witness, the existence of robust states
for which $W_{M}\geq0$ cannot be excluded. Below, we demonstrate a necessary and sufficient
criterion for robustness of Gaussian states, effectively determining the boundary
between robust and fragile states.

\section{Entanglement robustness: General Conditions\label{realrobust}}

In order to obtain clear-cut conditions for the robustness of entanglement,
we must employ a necessary and sufficient entanglement criterion.
By analyzing whether the subsystems remain entangled or become separable
upon attenuation, we will classify all bipartite Gaussian states.

\subsection{The PPT Criterion}

We find a convenient separability criterion in the requirement of positivity under partial 
transposition (PPT) of the density matrix for separable states~\cite{Peres,Horodecki}. An 
entangled state, on the other hand, will necessarily lead to a negative partially transposed 
density matrix, which is non-physical.

The partial transposition (PT) of the density operator is equivalent in the level of the Wigner function to the operation 
of time-reversal applied to a single subsystem. On the covariance matrix level, time-reversal 
is obtained by changing the sign of the momentum (for harmonic oscillators), or the sign 
of the phase quadrature of one mode (for electromagnetic fields), in this manner affecting the 
sign of its correlations~\cite{Simon}.

Physical validity is assessed using Eq.~(\ref{physicality}). The uncertainty relation can be 
recast into a more explicit form by expressing it in terms of the determinants of the covariance 
matrix and its submatrices as
\begin{equation}
1+\det V -2\det C-\sum_{i=1,2}\det A_{j}\geq 0.
\label{Physic}
\end{equation}
The PT operation modifies the sign of $\det C$, resulting in the following condition for entanglement~\cite{Simon}
\begin{equation}
W_{ppt}=1+\det V+2\det C-\sum_{i=1,2}\det A_{j}<0.
\label{Sep}
\end{equation}
Since all separable states fulfill $W_{ppt}\geq0$, $W_{ppt}$ is a sufficient entanglement 
witness. For Gaussian states, it is a necessary witness as well, and the equation $W_{ppt}=0$ 
traces a clear boundary in the space of bipartite Gaussian states, setting apart the subspaces of 
separable and entangled states.

It is convenient to recall here that the purities of Gaussian 
states are directly related to the determinant of the covariance 
matrices~\cite{Braunstein05}
\begin{eqnarray}
\mu & = & (\det V)^{-\frac{1}{2}},\\
\mu_{j} & = & (\det A_{j})^{-\frac{1}{2}},
\label{pursub}
\end{eqnarray}
so that the entanglement witness of Eq.~(\ref{Sep}) involves the total purity of the systems,
the purity of each subsystem, and the shared correlations.

\subsection{Covariance Matrix under Attenuation}

Applying the witness of Eq.~(\ref{Sep}) to the attenuated covariance matrix of Eq.~(\ref{attTrans}), one obtains
\begin{equation}
W_{ppt}'(T_1,T_2)=1+\det V'+2\det C'-\sum_{j=1,2}\det(A_i'),
\label{SepESD}
\end{equation}
from which $W_{ppt}'(T_1=1,T_2=1)=W_{ppt}$. From Eq.~(\ref{attTrans}), it follows that the individual 
submatrices transform as $C'=\sqrt{T_{1}T_{2}}C$ and $A_{j}'=T_{j}(A_{j}-I)+I$ under attenuations. 
The bilinear dependence of Eq.~(\ref{W_M}) on $T_1$ and $T_2$ which led to a constant sign of the 
witness is not expected here and robustness is not a general feature of bipartite entangled states.

In Appendix A, we derive an explicit transmittance-dependent form of $W_{ppt}'(T_1,T_2)$. We can factor 
out a term $T_1T_2$, which cannot change the sign of $W_{ppt}$. It assumes the form
\begin{equation}
W_{ppt}'(T_1,T_2) = T_1T_2W_R(T_1,T_2).
\label{Dfinal}
\end{equation}
The reduced witness $W_R$ preserves the sign of $W_{ppt}'$ (except for $T_1=T_2=0$, for which 
we know both modes are in their vacuum states and $W_{ppt}'=0$), maintaining only the relevant 
dependence on $T_1$ and $T_2$. It reads
\begin{equation}
W_R(T_{1},T_{2}) = T_{1}T_{2}\Gamma_{22}+T_{2}\Gamma_{12}+T_{1}\Gamma_{21}+\Gamma_{11}.
\label{WE}
\end{equation}
The expressions for the coefficients $\Gamma_{ij}$ in terms of the covariance matrix entries are given 
in Appendix A. We note that they are regarded as constants here, independent of $T_1$ and $T_2$. 

\begin{figure*}[!ht]
\includegraphics[width=17cm]{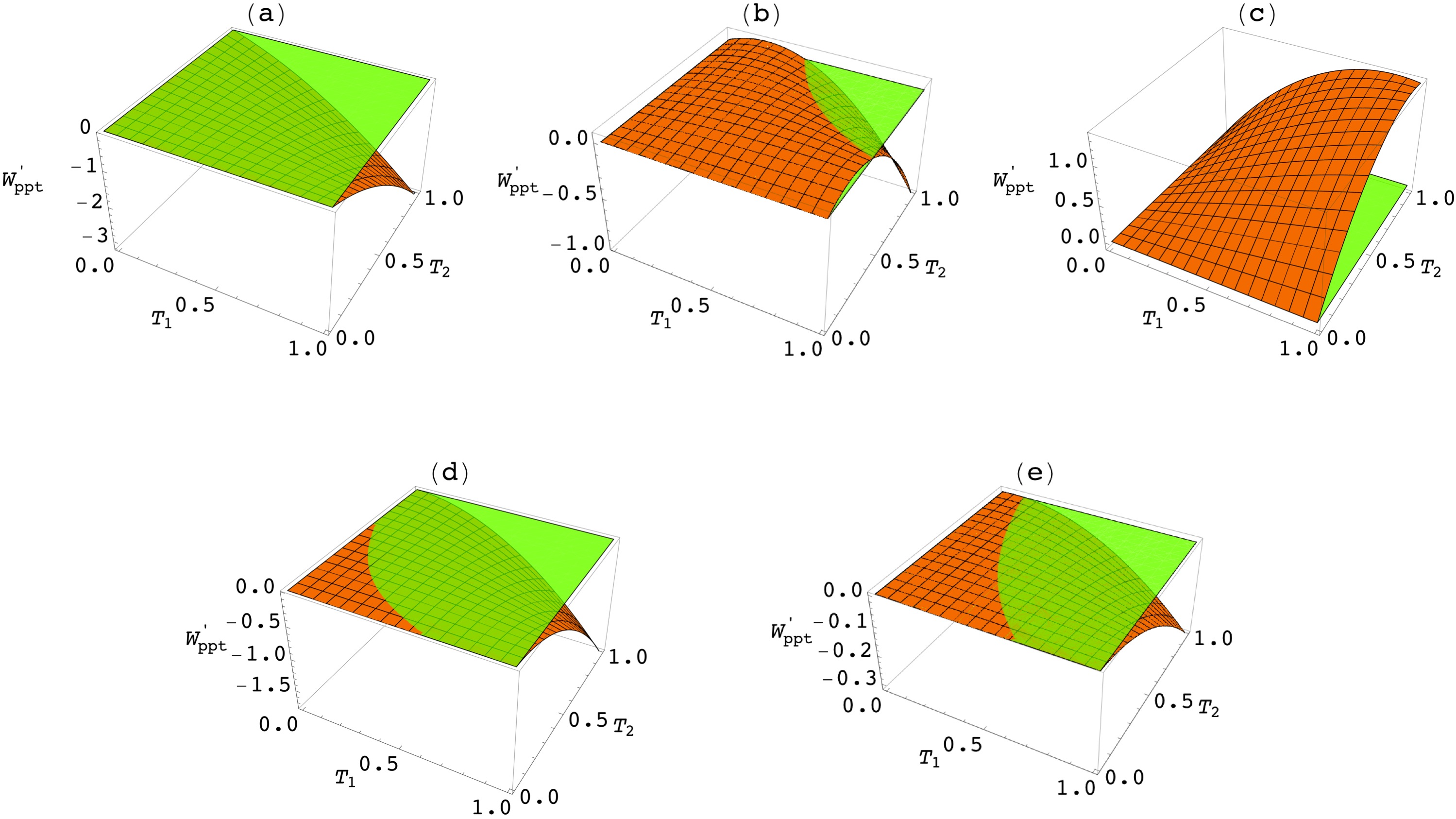}
\caption{(Color online) Possible behaviors of the PPT entanglement witness $W_{ppt}'$ under attenuation, as a 
function of the transmittances $T_1$ and $T_2$. (a) Fully robust entanglement. (b) Fragility for any combination of beam attenuations. (c) Separable state. (d) Single-channel partial robustness - 
either mode: the state is robust for any individual attenuation, but not for a combination of attenuations, 
such as equal attenuations.  (e) Single-channel partial robustness - specific mode, i.e., the state is 
robust when one mode is attenuated but presents ESD upon attenuation of the other mode.}
\label{simon3d}
\end{figure*}

The different dynamics of entanglement under losses appear in the witnesses $W_{ppt}'$ and $W_{R}$. 
Fig.~\ref{simon3d} depicts four entangled states (three of them fragile) plus a separable state 
under attenuation. The plots show $W_{ppt}'(T_1,T_2)$ based on the covariance matrix
\begin{eqnarray}
V=\left[\begin{array}{cccc}
\Delta^2 q_1 & 0 & c_q & 0\\
0 & \Delta^2 p_1 & 0 & c_p\\
c_q & 0 & \Delta^2 q_2 & 0\\
0 & c_p & 0 & \Delta^2 p_2
\end{array}\right],
\label{covsimples}
\end{eqnarray}
constructed from diagonal submatrices. This simple form of $V$, observed in the experiments of Ref.~\cite{Barbosa2010}, suffices to span all types of entanglement 
dynamics of Gaussian states. 

The curves of Fig.~\ref{simon3d}a--d were specifically obtained from
\begin{eqnarray}
V=\left[\begin{array}{cccc}
2.55 & 0 & c_q & 0\\
0 & 1.80 & 0 & -1.26\\
c_q & 0 & 2.55 & 0\\
0 & -1.26 & 0 & 1.80
\end{array}\right].
\label{covsimples2}
\end{eqnarray}
As the correlation $c_q$ is varied, different types of entanglement dynamics are observed. Modifying this parameter while keeping constant the other entries of the covariance matrix is equivalent to adding uncorrelated noise to the system (for instance, classical phonon noise dependent on the temperature of the non-linear crystal~\cite{Barbosa2010, pra09}). 
In Fig.~\ref{simon3d}a ($c_q=1.275$), a state violating the Duan criterion is {\it fully robust}, as expected. 
Disentanglement does not occur for finite losses imposed on any of the fields. In Fig.~\ref{simon3d}b, 
the choice $c_q=0.893$ characterizes a state for which ESD occurs for partial attenuation in a single channel 
(mode) or in both channels. This represents the most fragile class of states. In Fig.~\ref{simon3d}c ($c_q=0.3825$), 
the initial state is separable and it naturally remains separable throughout the whole region of attenuations.

A more subtle entanglement dynamics appears in Fig.~\ref{simon3d}d ($c_q=1.033$). The state is robust 
against any single-channel attenuation but may become separable if both modes are attenuated. Such a 
state would suffice as a resource for quantum communications involving single-channel losses.

If we consider a more general covariance matrix, with asymmetric modes, the system may be robust against 
losses on one mode, but not on the other. This is observed in Fig.~\ref{simon3d}e, where $W_{ppt}'$ is 
calculated for the covariance matrix
 \begin{eqnarray}
V=\left[\begin{array}{cccc}
2.55 & 0 & 0.653 & 0\\
0 & 1.80 & 0 & -0.797\\
0.653 & 0 & 1.62 & 0\\
0 & -0.797 & 0 & 1.32
\end{array}\right].
\label{covsimples3}
\end{eqnarray}
This particular covariance matrix is obtained from Eq.~(\ref{covsimples2}), with $c_q=1.033$, by imposing 
the attenuation $T_2=0.40$. Before this attenuation, the state was partially robust, as in Fig.~\ref{simon3d}d. 
It remains robust against losses on mode 2, but now disentanglement with respect to losses solely on mode 1 
may occur. This illustrates the fact that the new states produced upon attenuation become 
more fragile.  Since attenuation is a Gaussian operation, states cannot become more robust upon attenuation~\cite{distgauss1,distgauss2}. 

\subsection{Full Robustness}\label{FulRobSubsection}

We show here that fully robust states can be directly identified from the covariance matrix. 
In order to obtain the necessary condition, we note from Eq.~(\ref{WE}) that the entanglement dynamics 
close to complete attenuation is dominated by $\Gamma_{11}$. Thus, an initially entangled state $W_R(T_1=1,T_2=1)<0$ 
with $\Gamma_{11}>0$, must become separable for sufficiently large attenuation, from which we derive the witness
\begin{equation}
W_{full}=\Gamma_{11}=\sigma_{1}\sigma_{2}-\mathrm{tr}(C^{T}C)+2\det C.
\label{WRe}
\end{equation}
$W_{full}\leq 0$, provided $W_{ppt}<0$, supplies a simple, direct, and general condition for testing the entanglement robustness of 
bipartite Gaussian states.

The robustness cannot depend on the choice of local measurement basis for each mode since, as discussed in 
Appendix A, local rotations commute with the operation of losses. In other words, local passive operations, such as 
rotations and phase shifts, do not change the robustness. By using local rotations 
to diagonalize the correlation matrix $C$, we obtain
\begin{equation}
W_{full}^{(\mathit{D})}=\sigma_{1}\sigma_{2}-(c_{p}-c_{q})^{2}\leq0,
\label{WMgeral}
\end{equation}
which coincides with $W_M$, of Eq.~(\ref{W_M}). Thus, the Duan criterion in the simple form of Eq.~(\ref{W_M}) 
is a particular case of Eq.~(\ref{WRe}) when the correlation submatrix is diagonal. For Gaussian states given by 
covariance matrices with diagonal correlation submatrix, $W_{M}$ is a necessary and sufficient witness for robust 
entanglement, but only sufficient otherwise. 

\subsection{Partial Robustness}\label{ParRobSubsection}

As seen in Fig.~\ref{simon3d}, there exist states which can be robust against single-channel 
losses, yet disentangle for finite losses split among two channels. Similar to the procedure in the 
previous section, we will define witnesses capable of identifying partial robustness. 

Let us consider the case $T_2=1$ for definiteness. The attenuated witness of Eq.~(\ref{WE}) becomes
\begin{equation}
W_R(T_1,T_2=1) = (W_{ppt}-W_{1})T_1+W_{1},
\label{WitRob1}
\end{equation}
where
\begin{equation}
W_{1} = W_{full}+\Gamma_{21}
\label{WitRob2}
\end{equation}
(see Appendix A for the expression of $\Gamma_{21}$). The analysis of $W_{1}$ follows the same lines 
used in the case of fully robust states, with the simplification that the witness depends linearly on the 
attenuation. Thus, there is only one possible path cutting the plane $W_R(T_1,T_2=1)=0$.
The fraction of transmitted light for which ESD occurs is
\begin{eqnarray}
T_1^c & = & \frac{W_{1}}{W_{1}-W_{ppt}}.
\label{TCrit}
\end{eqnarray}
From $W_{ppt}<0$, it follows that $0<W_{1}<W_{1}-W_{ppt}$, to assure that $T_1^c$ exists as a 
meaningful physical quantity ($0<T_1^c<1$) whenever $W_{1}>0$.

Therefore, an entangled state satisfying $W_{1}\leq 0$ is robust against losses in channel 1, and 
$W_1$ is the witness for this type of robustness. The corresponding analysis regarding attenuations 
on the subsystem 2 yields the witness
\begin{equation}
W_{2}=W_{full}+\Gamma_{12},
\label{W2}
\end{equation}
with the same properties of $W_{1}$. A relation analogous to Eq.~(\ref{TCrit}) holds for $T_2^c$. 
Both witnesses are invariant under local rotations, as expected.

\section{Robustness Classes\label{class}}

Based on the different dynamics of entanglement of Fig.~(\ref{simon3d}), we 
propose a classification of bipartite entangled states according to their resilience 
to losses. We take guidance in the sign of the reduced witness $W_R(T_{1},T_{2})$, which 
is a hyperbolic paraboloid surface. The contour defined by the condition $W_R(T_{1},T_{2})=0$ 
provides a complete description of the entanglement dynamics in terms of $\Gamma_{ij}$. As depicted 
in Fig.~\ref{simon3d}, there are three relevant situations. Bipartite entangled Gaussian states can be 
assigned to the following different classes:
\renewcommand{\theenumi}{(\roman{enumi})}
\begin{enumerate}
\item {\it Fully robust states} remain entangled for any partial attenuation: $W_R(T_1,T_2)<0, \forall T_{1,2}$.
\item {\it Partially robust states: (a) symmetric} -- remain entangled against losses on a single mode, but may 
disentangle for combinations of partial attenuations on both modes:
     $W_R(T_1,T_2=1)<0, \forall T_{1}$, and $W_R(T_1=1,T_2)<0, \forall T_{2}$.
 {\it (b) asymmetric} -- remain entangled against losses on a specific mode, but may disentangle for partial losses 
 on the other mode: either $W_R(T_1,T_2=1)<0, \forall T_{1}$, or $W_R(T_1=1,T_2)<0, \forall T_{2}$.
\item {\it Fragile states} disentangle for partial attenuation on any mode or combinations of partial attenuations on both modes.
\end{enumerate}
For a complete classification of all bipartite Gaussian states, one should include the separable states. 

With the witnesses previously defined, we have necessary criteria to assess the robustness of all bipartite 
Gaussian states. A state will be robust with respect to losses imposed on subsystem $1$ if
\begin{equation}
W_{1}\leq 0 \; .
\label{ParEntWit}
\end{equation}
Likewise, robustness to losses on subsystem $2$ is given by
\begin{equation}
W_{2}\leq 0.
\end{equation}
States will be partially robust if at least one of $W_1$ or $W_2$ is negative or even if both 
are negative simultaneously (partially robust -- symmetric). Only if  $W_R(T_1,T_2)<0, \forall T_{1,2}$ 
will we have full robustness. 

As mentioned above, this classification is of practical interest. Several quantum communication 
protocols using continuous variables can be realized by one of the parties (Alice) locally producing 
the entangled state and sending only one mode to a remote location. The other party (Bob) then 
performs operations on his part of the state, according to instructions sent by Alice through a 
classical channel. The success of these communication schemes strongly depends on the 
losses that the subsystem of Bob may undergo, which could be detected by an eavesdropper (Eve). In 
this situation, Alice must produce entangled states that are at least partially robust in order to avoid problems 
with signal degradation. It may not be necessary for her to produce fully robust states: partially robust entangled 
states may suffice for successful quantum communication protocols.

\section{Particular Cases \label{examples}} 

In the preceding analysis we have found precise conditions to determine 
whether or not bipartite Gaussian entangled states are robust against losses. 
Given the practical interest of such states as resources for quantum communication 
protocols, we examine here particular Gaussian states that fall within the 
classification scheme proposed above. One might think that it should suffice to 
generate pure states with a large amount squeezing in order to have robust 
entanglement. We begin by providing a specific example of a pure strongly 
squeezed state, which is only partially robust. We then examine different forms 
of the covariance matrix, in order to map out the different possibilities.

\subsection{Pure and highly squeezed states with only partial robustness} 



In most experiments, Gaussian bipartite entanglement is witnessed by a 
violation of the simplified Duan inequality of Eq.~(\ref{Duan}). Typically, 
this is done by combining highly squeezed individual modes on a beam splitter. 
This method allows the creation of arbitrarily strong entanglement in the sense 
that quantum information protocols such as teleportation could in principle be 
realized with perfect fidelity in the limit of an EPR state. 

If such a state is contaminated by uncorrelated classical noise (e.g. from 
Brillouin scattering in an optical fiber~\cite{elser}), it may then become subject to 
disentanglement from losses. Even states which are pure may be subject 
to disentanglement in a dual-channel scenario. We present below the 
covariance matrix for a pure state with these characteristics: 

\begin{equation}
V=\left(\begin{array}{cccc}
52.5 & 0 & -47.5 & 0\\
0 & 0.105 & 0 & 0.095\\
-47.5 & 0 & 52.5 & 0\\
0 & 0.095 & 0 & 0.105\end{array}\right).
\label{HighlySqueezed}
\end{equation} 

\begin{figure}[ht] 
\includegraphics[width=8.25 cm]{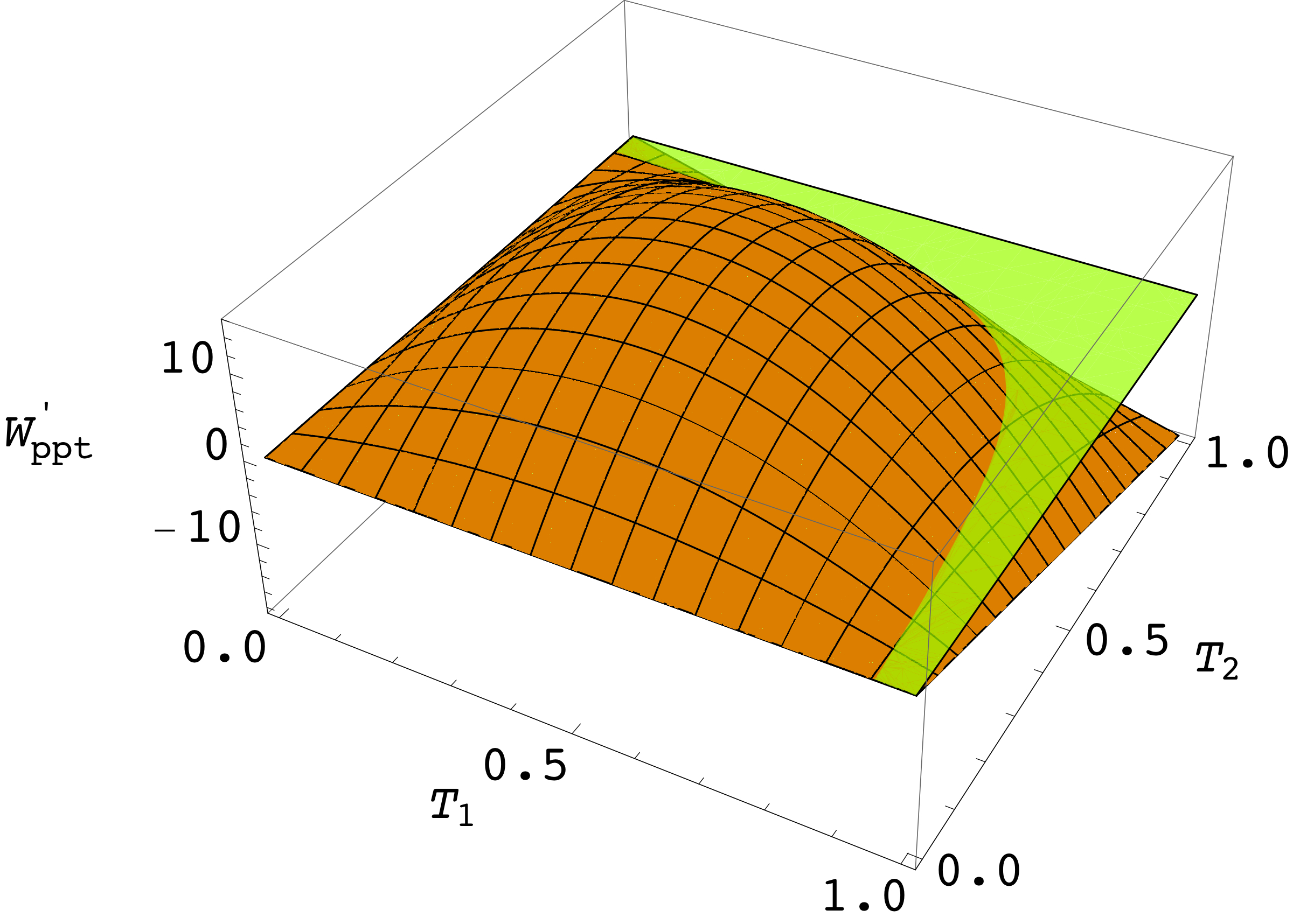}
\caption{(Color online) Entanglement as function of losses for the covariance matrix given by 
Eq.~(\ref{HighlySqueezed}). Disentanglement may occur only for combined losses on 
both modes. In this example, the symplectic eigenvalue \cite{Simon} is only 0.22 for the initial 
state.}
\label{fig:Hisqz}
\end{figure}

This state has a very small symplectic eigenvalue, indicating very 
strong entanglement~\cite{Vidal}. As can be observed in Fig.~\ref{fig:Hisqz}, the state is 
partially robust: losses on any single channel do not lead to disentanglement, 
while ESD will occur for combined losses in both channels. 

Let us now examine different symmetries of the covariance matrix and their 
implications on the entanglement dynamics.

\subsection{Symmetric Modes and Quadratures  - Fully Robust States}
 
We begin by examining completely symmetric modes, for which
$\Delta^2\hat p_1=\Delta^2\hat q_1=\Delta^2\hat p_2=\Delta^2\hat q_2=s$
and $\langle\delta\hat p_1\delta\hat p_2\rangle=\langle\delta\hat q_1\delta\hat q_2\rangle=c$,
and $\langle\delta\hat p_j \delta\hat q_{j'}\rangle=0$. The covariance matrix has the form
\begin{equation}
V=\left(\begin{array}{cccc}
s & 0 & c & 0\\
0 & s & 0 & -c\\
c & 0 & s & 0\\
0 & -c & 0 & s\end{array}\right).
\label{fully-sym}
\end{equation}
Such states can be generated, for instance, by the interference of (symmetric) squeezed states on a 
balanced beam splitter (entangled squeezed states)~\cite{KimbleFurusawa, Leuchs}. In 
this case one has $s=\nu\cosh2r$ and $c=\nu\sinh 2r$, where $r$ is the squeezing parameter and 
$\nu\geq1$ accounts for an eventual thermal mixedness, representing a correlated classical noise 
between the systems.

Entanglement and robustness witnesses are thus
\begin{equation}
W_{ppt}=(s^2-c^2+1)^2-4s^2
\label{simon-fully-sym}
\end{equation}
and
\begin{equation}
W_{full}=4[(s-1)^{2}-c^{2}]=4(s^{2}-c^{2}+1-2s),
\label{esr-fully-sym}
\end{equation}
from which one directly sees that $W_{ppt}<0$ and $W_{full}<0$ lead to the same condition ($s-1-|c|<0$). 
Therefore, entangled states with symmetry between the two modes and the two quadratures are fully 
robust. The lack of ESD in these systems indicates that strong symmetries lead to entanglement robustness, 
even when classical noise is present, as long as it is correlated.

The highly symmetric covariance matrices of Eq.~(\ref{fully-sym}) are a particular case of the standard form II of 
Ref.~\cite{DGCZ}. For these, the Duan criterion is equivalent to the PPT criterion, which then entails full 
robustness for all entangled states with covariance matrices in standard form II. Moreover, since any state
can be brought to standard form II by local squeezing and quadrature rotations without changing its 
entanglement~\cite{DGCZ}, any fragile state can be made robust by suitable local unitary operations. The 
converse is also true: local squeezing can transform robust states into fragile ones without changing the 
entanglement. For instance, if one applies a gate that makes use of local squeezing to a given robust 
entangled state, it can become fragile and undergo disentanglement upon transmission. Local squeezing 
is one of the important steps in an implementation of a C-NOT (or QND) gate with continuous 
variables~\cite{Yoshikawa2008}.

\subsection{Symmetric Modes and Asymmetric Quadratures}

More general covariance matrices are necessary in order to observe disentanglement. States which are 
symmetric on both modes but asymmetric on the quantum statistics of the quadratures have been recently 
observed to present ESD~\cite{Barbosa2010}. The system under investigation consisted of the twin light 
beams produced by an optical parametric oscillator, described by a covariance matrix of the form
\begin{equation}
V=\left(\begin{array}{cccc}
\Delta^2 q & 0 & c_{q} & 0\\
0 & \Delta^2 p & 0 & c_{p}\\
c_{q} & 0 & \Delta^2 q & 0\\
0 & c_{p} & 0 & \Delta^2 p
\end{array}\right).
\label{sym-mode}
\end{equation}
The entanglement and robustness witnesses read
\begin{eqnarray}
W_{ppt}=&\left[(\Delta^2 p)^{2} -  c_{p}^{2}\right]\left[(\Delta^2 q)^{2}-c_{q}^{2}\right]\nonumber\\
&\qquad \qquad - 2\Delta^2 p\,\Delta^2 q+2c_{p}c_{q}+1
\label{simon-symmetric}
\end{eqnarray}
and
\begin{equation}
W_{full}=(\Delta^2 p+\Delta^2 q-2)^{2}-(c_{q}-c_{p})^{2}.
\label{esd-symmetric}
\end{equation}
In this situation, the subsystems have equal purities ($\mu_S=1/\sqrt{\Delta^2 p\Delta^2 q}$). The quadrature 
variances and correlations are constrained by $(\Delta^2 p)^2-c_p^2\geq0$ and 
$(\Delta^2 q)^{2}-c_q^2\geq0$. We introduce the normalized correlations $\bar C_p=c_p/\Delta^2 p$ 
and $\bar C_q=c_q/\Delta^2 q$ for simplicity. They are bounded by $-1\leq\bar C_j\leq1$. 
These parameters suffice to describe any state with the form of Eq.~(\ref{sym-mode}).

\begin{figure}[ht]
\includegraphics[width=8.25 cm]{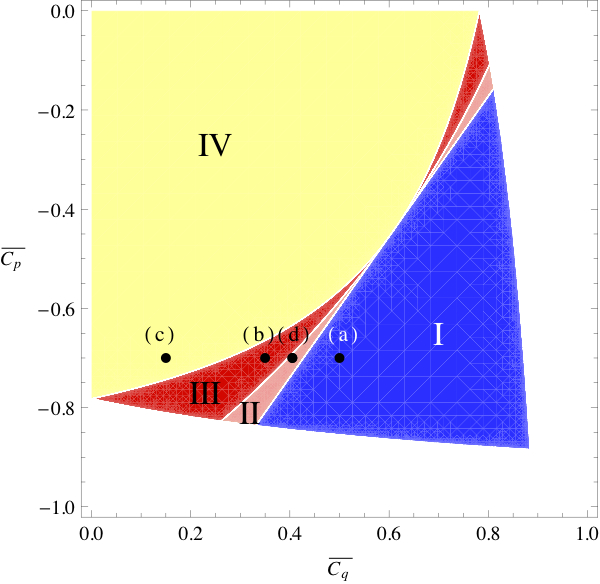}
\caption{ (Color online) The space of states with covariance matrices of the form of Eq.~(\ref{sym-mode}) is plotted as 
a function of the normalized correlations $\bar{C}_{p}$ and $\bar{C}_{q}$. Separable states lie in the region \textbf{IV} 
; fully robust states are comprised within the region \textbf{I}; partially robust states are in the region \textbf{II}, 
and fragile states are in the region \textbf{III}. Points outside of these regions do not correspond to physical states. 
Here we use $\Delta^2 p=1.80$ and $\Delta^2 q=2.55$. The points included represent the states in Fig.~\ref{simon3d}a--d. }
\label{symmetric-beams}
\end{figure}

In Fig.~\ref{symmetric-beams} the robustness condition is mapped in terms of the correlations for a fixed purity 
$\mu_S= 0.626$, showing the regions corresponding to different robustness classes. Fully robust state (a) 
falls within the \textbf{I} region in Fig.~\ref{symmetric-beams}, while the separable state (c) is located in the \textbf{IV} 
region. Within the intermediate region, two different types of \textit{fragile} states are present. State (d) is partially robust, 
lying close to the boundary to robust states. State (b) shows ESD for partial losses in general, lying close to the 
boundary to separable states.

Alternatively, following the treatment described in Ref.~\cite{Barbosa2010}, the covariance matrix 
of Eq.~(\ref{sym-mode}) can be parametrized in terms of the physically familiar EPR-type operators,
\begin{equation}
\hat{p}_{\pm}=\frac{1}{\sqrt{2}}(\hat{p}_{1}\pm\hat{p}_{2})
\label{pmm}
\end{equation}
and
\begin{equation}
\hat{q}_{\pm}=\frac{1}{\sqrt{2}}(\hat{q}_{1}\pm\hat{q}_{2}).
\label{qmm}
\end{equation}

Entanglement can be directly observed from the product of squeezed variances of the proper pair 
of EPR operators, $(\hat p_-,\hat q_+)$ or $(\hat p_+,\hat q_-)$. Additionally, the entanglement and robustness criteria of symmetric two-mode systems of Eqs.~(\ref{simon-symmetric})--(\ref{esd-symmetric}) 
can be written in the simpler forms,
\begin{eqnarray}
W_{ppt}&=&W_{prod}\overline{W}_{prod},
\label{simon-mmm}\\
W_{full}&=&W_{sum}\overline{W}_{sum},
\label{esd-mmm}
\end{eqnarray}
where
\begin{eqnarray*}
W_{sum}&=&\Delta^2\hat{p}_-+\Delta^2\hat{q}_+-2,\\
\overline W_{sum}&=&\Delta^2\hat{p}_++\Delta^2\hat{q}_--2,\\
W_{prod}&=&\Delta^2\hat{p}_-\Delta^2\hat{q}_+-1,\\
\overline W_{prod}&=&\Delta^2\hat{p}_+\Delta^2\hat{q}_--1.
\end{eqnarray*}

The distinction between robust and partially robust entanglement is clearly illustrated with symmetric modes. 
Considering attenuation solely on mode 1 (entirely equivalent to attenuation on mode 2, given the 
symmetry), the condition for partial robustness of Eq.~(\ref{ParEntWit}) yields
\begin{equation}
\label{partial-esd-mmm}
W_{1}=W_{sum}\overline{W}_{prod}+W_{prod}\overline{W}_{sum}.
\end{equation}
The condition $W_{1}=0$ defines the border between partial robustness and fragility. Since a state must 
be initially entangled in order to disentangle, obviously
\begin{equation}\label{LogCond}
W_{full}<0\Longrightarrow W_{ppt}<0.
\end{equation}
Given the commutation relations between $\hat{p}$ and $\hat{q}$, $W_{prod}$ and $\overline{W}_{prod}$ 
(or $W_{sum}$ and $\overline{W}_{sum}$) cannot be simultaneously negative. 
In this context, the condition of Eq.~(\ref{LogCond}) can be restated as
\begin{eqnarray}
W_{sum}<0 & \Longrightarrow & W_{prod}<0 \;\mbox{or} \nonumber \\ 
\overline{W}_{sum}<0 & \Longrightarrow & \overline{W}_{prod}<0.
\end{eqnarray}
For $W_{1}=0$,
\begin{equation}
\label{partial-esd-mmm2}
W_{sum}\overline{W}_{prod}=-W_{prod}\overline{W}_{sum}.
\end{equation}
This equation holds only if $W_{prod}<0$ and $W_{sum}>0$ (or $\overline{W}_{prod}<0$ and $\overline{W}_{sum}>0$). 
Thus $W_{1}=0$ lies between the curves $W_{ppt}=0$ and $W_{full}=0$.

\begin{figure}[ht]
\includegraphics[width=8.25 cm]{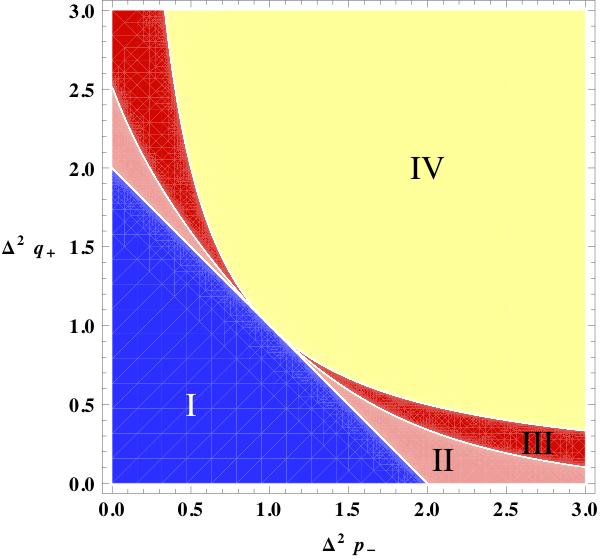}
\caption{(Color online) The space of symmetric two-mode states is plotted as a function of the EPR variances 
$\Delta^2\hat{q}_+$ and $\Delta^2\hat{p}_-$, normalized to the standard quantum limit (SQL). Separable states lie in the region \textbf{IV}; fully robust entangled states 
are within the region \textbf{I}; partially robust states are in the region \textbf{II} and fragile states are in the region
\textbf{III}. The partial purities are $\mu_{-}=0.7267$ and $\mu_{+}=0.4529$. 
}
\label{symmetric-beams-mmm}
\end{figure}

A plot of the state space in terms of these EPR variables is presented in Fig.~\ref{symmetric-beams-mmm}. Fixed 
values for the partial purities, $\mu_+=1/\sqrt{\Delta^2\hat{p}_+\Delta^2\hat{q}_+}$ and 
$\mu_-=1/\sqrt{\Delta^2\hat{p}_-\Delta^2\hat{q}_-}$, are assumed, so that we can write the entanglement and 
robustness conditions in terms of $\Delta^2\hat{p}_-$ and $\Delta^2\hat{q}_{+}$. The observation of ESD reported 
in Ref.~\cite{Barbosa2010} was obtained for partially robust states lying in the region delimited by the conditions 
$W_{sum}>0$ and $W_1<0$.

\subsection{System in Standard Form I}

The last case we consider is a covariance matrix in the standard form~I~\cite{Simon,DGCZ}.
It represents two different modes with symmetric quadratures,
\begin{equation}
V=\left(\begin{array}{cccc}
s & 0 & c_{q} & 0\\
0 & s & 0 & c_{p}\\
c_{q} & 0 & t & 0\\
0 & c_{p} & 0 & t\end{array}\right).
\end{equation}
The entanglement and full robustness witnesses read
\begin{equation}
W_{ppt}=(st-c_{q}^{2})(st-c_{p}^{2})-s^{2}-t^{2}+2c_{q}c_{p}+1
\end{equation}
and
\begin{equation}
W_{full}=4(s-1)(t-1)-(c_{q}-c_{p})^{2}.
\end{equation}
The subsystems have purities $\mu_{1}=s^{-1}$ and $\mu_{2}=t^{-1}$. We define the normalized 
correlations $\bar{c}_{j}=c_{j}/\sqrt{st}=c_{j}\sqrt{\mu_{1}\mu_{2}}$ as before.


A covariance matrix in standard form I also presents ESD for certain parameters, 
spanning all three classes of states described above. 
Owing to the symmetry in the covariance matrix, ESD in such a system does not occur 
for symmetric correlations, $\bar{c}_{q}=-\bar{c}_{p}$, independently 
of the purities $\mu_{1}$ and $\mu_{2}$. 
\section{Conclusion \label{conclusion}}

We have addressed in this paper the issue of entanglement in the open-system dynamics of 
continuous-variable (CV) systems. Entanglement is a crucial albeit fragile resource for quantum information 
protocols. Understanding its behavior in open systems is very important for future practical applications. 

Our analysis is carried out for the simplest possible situation in the CV setting: bipartite Gaussian states 
under linear losses. The general study undertaken here 
was motivated by the experimental results presented in~\cite{CoelhoScience,Barbosa2010}. 

Starting from necessary and sufficient entanglement criteria, we 
derived necessary and sufficient {\em robustness} criteria, which enable us to classify these states 
with respect to their entanglement resilience under losses. Having in mind realistic communications 
scenarios, we present a robustness classification: states may be fully robust, partially robust, or fragile. 
For instance, if one generates an entangled 
state for which only one mode will propagate in a lossy quantum channel (single-channel losses), 
the conditions derived for partially robust states apply. Such partial robustness would be the minimum 
resource required for single-channel robust quantum communications. 

On the other extreme, EPR states, for which quantum correlations appear in collective operators of
both quadratures, are the best desirable quantum resource. Their entanglement is resilient to
any combination of losses acting on both modes, only disappearing when the state suffers total loss. 
However, a rather likely deviation from such states could already be catastrophic for entanglement: if
a moderate amount of uncorrelated noise (e.g. thermal noise) is introduced in the EPR-type collective operators for one 
quadrature, even when the other quadrature remains untouched and is perfectly squeezed, entanglement 
can be lost for partial attenuation. This offers a clue to the main ingredients leading to ESD in 
bipartite Gaussian states. An appealing example is given by the OPO operating above threshold. 
The usual theoretical analysis leads to symmetric modes, with asymmetric quadratures, but no 
uncorrelated classical noise. Thus, the OPO is predicted to generate fully robust entangled states. 
However, uncorrelated thermal noise originating in the non-linear crystal couples into the two modes~\cite{pra09}, 
leading to ESD~\cite{Barbosa2010}. 

We have also found that such noise does not necessarily have to imply mixedness. 
Even for pure states, the lack of correlation between modes increases the state's fragility. Robustness 
is thus achieved not only for high levels of entanglement between CV systems, but also symmetry 
in the form of quantum correlations is desirable. This point was illustrated by our study of mathematical 
examples of Gaussian states, for which symmetry implied robustness in spite of mixedness. 
We also point out that robustness can be obtained, in principle, for any entangled state by local unitary 
operations, such as squeezing and quadrature rotations. However, these operations are not 
always simple to implement in an experiment. 

As an outlook, we should keep in mind that scalability is one of the main goals in 
quantum information research at present. As larger and more complex systems are 
envisioned for the implementation of useful protocols, higher orders of entanglement 
will be required. Disentanglement for partial losses was experimentally observed in 
the context of a tripartite system~\cite{CoelhoScience}. An understanding of entanglement 
resilience for higher-order systems will be important. The methods and analyses developed 
here constitute the starting point for such investigations. 

\acknowledgements

This work was supported by the Conselho Nacional de Desenvolvimento Cient\'\i{}fico e 
Tecnol\'ogico (CNPq) and the Funda\c{c}\~ao de Amparo \`a Pesquisa do Estado S\~ao Paulo 
(FAPESP). KNC and ASV acknowledge support from the AvH Foundation.

\appendix

\section{Attenuated Witness \label{appA}}

We would like to obtain an explicit expression for $W_{ppt}'(T_1,T_2)$ in terms of the physical parameters 
of the bipartite system (variances and correlations). We note that the procedure cannot be directly realized 
by first bringing $V'$ (or $V$) to a standard form and then applying the attenuation, since local symplectic 
operations  $S\in\mathrm{Sp}(2,\Re)\oplus\mathrm{Sp}(2,\Re)$ normally do not commute with the attenuation operation, 
$\mathcal{L}(SVS^{T})\neq S\mathcal{L}(V)S^{T}$~\cite{Eisert05, key-1}. Consequently, invariant quantities 
under global and local symplectic transformations are not necessarily conserved by attenuations, such as the 
global and local purities. On the other hand, $S\mathcal{L}(V)S^{T}=\mathcal{L}(SVS^{T})$ is satisfied only if 
$SS^{T}=I$, i.e. $S$ must be a local phase space rotation,  $S\in\mathrm{SO}(2,\Re)\oplus\mathrm{SO}(2,\Re)$. 
Therefore, a criterion for entanglement robustness should depend solely on local rotational invariants.

We derive the explicit behavior of the witness $W_{ppt}'$ under attenuation.
Writing the PPT separability criterion in terms of the symplectic
invariants~\cite{Simon}, we obtain 

\begin{eqnarray}
W_{ppt} & = & 1+\det V+2\det C-\sum_{j=1,2}\det A_{j},\label{SepApp}\\
\det V & = & \det A_{1}\det A_{2}+\det C^{2}-\Lambda_{4},\label{DetV}\\
\Lambda_{4} & = & \mathrm{tr}(A_{1}JCJA_{2}JC^{T}J).\label{I4}\end{eqnarray}
After attenuation, the matrices $A_{1}$, $A_{2}$, and $C$ become
\begin{eqnarray}
C' & = & \sqrt{T_{1}T_{2}}C,\label{Ctrans}\\
A_{i}' & = & T_{i}(A_{i}-I)+I,\label{Atrans}\end{eqnarray}

To derive Eq.~(\ref{WE}), we express the symplectic invariants in terms of quantities presenting 
similar behavior. Two such quantities are obtained from Eq.~(\ref{attTrans})
and Eq.~(\ref{Ctrans}),
\begin{eqnarray}
\det(V'-I) & = & T_{1}^{2}T_{2}^{2}\det(V-I),\label{DetVprime}\\
\det C' & = & T_{1}T_{2}\det C.\label{DetC}
\end{eqnarray}
Since for any $2\times2$ matrix $M$ the following expressions are
valid, \begin{eqnarray}
\det(M-I) & = & \det M-\mathrm{tr}M+1,\label{DetTr}\\
\mathrm{tr}(M-I) & = & \mathrm{tr}(M)-2,\label{TRTr}\end{eqnarray}
 one obtains
 \begin{eqnarray}
\varpi_{j}'-\sigma_{j}' & = & T_{j}^{2}(\varpi_{j}-\sigma_{j}),\label{DetA}\\
\sigma_{j}' & = & T_{j}\sigma_{j},\label{Trace}
\end{eqnarray}
where $\sigma_{i}=\mathrm{tr}A_{i}-2$, and $\varpi_{i}=\det A_{i}-1$ 
is the deviation from a pure state (impurity), which is zero for a pure state 
and positive for any mixed state.

Applying Eq.~(\ref{DetTr}) to $\det(V-I)$, we find quantities
which scale polynomially on the beam attenuations,
\begin{eqnarray}
\det V & = & \det(V-I)+\eta,\label{Vprime2}\\
\eta & = & \sigma_{1}(\varpi_{2}-\sigma_{2})+\sigma_{2}(\varpi_{1}-\sigma_{1})+\sigma_{1}\sigma_{2}\\
 & + & \det(A_{1})+\det(A_{2})+\Lambda_{1}+\Lambda_{2}-\Lambda_{C}-1\label{eta}\nonumber \\
\Lambda_{1} & = & \mathrm{tr}(C^{T}J(A_{1}-I)JC),\nonumber \\
\Lambda_{2} & = & \mathrm{tr}(CJ(A_{2}-I)JC^{T}),\nonumber \\
\Lambda_{C} & = & \mathrm{tr}(C^{T}C)
\end{eqnarray}
 where the last three quantities scale as
 \begin{eqnarray}
\Lambda_{1}'=T_{1}^{2}T_{2}\Lambda_{1}\; & , & \;\Lambda_{2}'=T_{1}T_{2}^{2}\Lambda_{2},\label{Tr1}\\
\Lambda_{C}' & = & T_{1}T_{2}\Lambda_{C}.\label{TrC}
\end{eqnarray}

Substituting Eq.~(\ref{Vprime2}) in Eq.~(\ref{SepApp}) and applying
the attenuation operation, we arrive at
\begin{eqnarray}
W_{ppt}'(T_{1},T_{2}) & = & \sum_{i,j=1,2}T_{1}^{i}T_{2}^{j}\Gamma_{ij},\,\mbox{with}\label{Dfinal-other}\\
\Gamma_{22} & = & \det(V-I)=\det(V)-\eta,\nonumber \\
\Gamma_{12} & = & \sigma_{1}(\varpi_{2}-\sigma_{2})+\Lambda_{2},\nonumber \\
\Gamma_{21} & = & \sigma_{2}(\varpi_{1}-\sigma_{1})+\Lambda_{1},\nonumber \\
\Gamma_{11} & = & \sigma_{1}\sigma_{2}-\Lambda_{C}+2\det(C),\nonumber
\end{eqnarray}
The function $W_{ppt}'$ describes the dynamics of all bipartite Gaussian states under losses.



\begin{thebibliography}{99}

\bibitem{Eberly}
T. Yu and J. H. Eberly,
Science \textbf{323}, 598 (2009), and references therein.

\bibitem{Davidovich}
M. P. Almeida, F. de Melo, M. Hor-Meyll, A. Salles, S. P. Walborn, P. H. Souto Ribeiro, and L. Davidovich,
Science \textbf{316}, 579 (2007).

\bibitem{CoelhoScience}
A. S. Coelho, F. A. S. Barbosa, K. N. Cassemiro, A. S. Villar, M. Martinelli, and P. Nussenzveig,
Science \textbf{326}, 823 (2009).

\bibitem{Barbosa2010}
F. A. S. Barbosa, A. S. Coelho, A. J. de Faria, K. N. Cassemiro, A. S. Villar, P. Nussenzveig, and M. Martinelli,
Nature Photon. {\bf 4}, 858 (2010).

\bibitem{braunsteinvanloockRMP}
S. L. Braunstein and P. van Loock, 
Rev. Mod. Phys. {\bf 77}, 513 (2005).

\bibitem{Simon}
R. Simon,
Phys. Rev. Lett. \textbf{84}, 2726 (2000).

\bibitem{wernerwolf}
R. F. Werner and M. M. Wolf, 
Phys Rev. Lett. {\bf 86}, 3658 (2001).

\bibitem{wolfeisertplenio2003} 
M. M. Wolf, J. Eisert, and M. B. Plenio, 
Phys. Rev. Lett. {\bf 90}, 047904 (2003). 

\bibitem{Simon94}
R. Simon, N. Mukunda, and B. Dutta,
Phys. Rev. A \textbf{49}, 1567 (1994).

\bibitem{DGCZ}
Lu-Ming Duan, G. Giedke, J. I. Cirac, and P. Zoller,
Phys. Rev. Lett. \textbf{84}, 2722 (2000).

\bibitem{Holevo01}
A. S. Holevo and R. F. Werner,
Phys Rev. A \textbf{63}, 032312 (2001).

\bibitem{Eisert05}
J. Eisert and M. M. Wolf,
in \textit{Quantum Information with Continuous Variables of Atoms and Light}, N. J. Cerf, G. Leuchs,
and E. S. Polzik, Eds., p. 23-42
(Imperial College Press, London, 2007); quant-ph/0505151 (2005).

\bibitem{EPR}
A. Einstein, B. Podolsky, and N. Rosen,
Phys. Rev. \textbf{47}, 777 (1935).

\bibitem{Bohr}
N. Bohr,
Phys. Rev. \textbf{48}, 696 (1935).

\bibitem{Leuchs}
Ch. Silberhorn, P. K. Lam, O. Weiss, F. Konig, N. Korolkova, and G. Leuchs,
Phys. Rev. Lett. \textbf{86}, 4267 (2001).

\bibitem{BowenPRL2003} W. P. Bowen, R. Schnabel, P. K. Lam, and T. C. Ralph, 
Phys. Rev. Lett. \textbf{90}, 043601 (2003).

\bibitem{KimbleFurusawa}
A. Furusawa {\it et al.},
Science \textbf{23}, 706 (1998).

\bibitem{VillarPRL}
A. S. Villar, L. S. Cruz, K. N. Cassemiro, M. Martinelli, and P. Nussenzveig, 
Phys. Rev. Lett. \textbf{95}, 243603 (2005).

\bibitem{Peres}
A. Peres,
Phys. Rev. Lett. \textbf{77}, 1413 (1996).

\bibitem{Horodecki}
M. Horodecki, P. Horodecki, and R. Horodecki,
Phys. Lett. A \textbf{223}, 1 (1996).

\bibitem{Braunstein05}
S. L. Braunstein and P. van Loock,
Rev. Mod. Phys. \textbf{77}, 513 (2005).

\bibitem{pra09} J. E. S. C\'esar, A. S. Coelho, K. N. Cassemiro, A. S. Villar, M. Lassen, P. Nussenzveig, and M. Martinelli,  
Phys. Rev. A {\bf 79}, 063816 (2009).

\bibitem{distgauss1} J. Eisert, S. Scheel and M. B. Plenio, 
Phys. Rev. Lett. \textbf{89}, 137903 (2002).

\bibitem{distgauss2} G. Giedke and J. I. Cirac, 
Phys. Rev. A \textbf{66}, 032316 (2002).

\bibitem{Vidal}
	G. Vidal and R. F. Werner,
	Phys. Rev. A \textbf{65}, 032314 (2002).

\bibitem{Yoshikawa2008}
  J. I. Yoshikawa, Y. Miwa, A. Huck, U. L. Andersen, P. van Loock, and A. Furusawa
  Phys. Rev. Lett. \textbf{101}, 250501 (2008).

\bibitem{elser}
D. Elser, U. L. Andersen, A. Korn, O. Gl\"ockl, S. Lorenz, Ch. Marquardt, and G. Leuchs,
Phys. Rev. Lett. {\bf 97}, 133901 (2006).

\bibitem{key-1}
Arvind, B. Dutta, N. Mukunda, and R. Simon,
Pramana {\bf 45}, 471 (1995); quant-ph/9509002.

\end{thebibliography}
\end{document}